\documentclass[journal]{IEEEtran}
\ifCLASSINFOpdf
\else
\fi

\usepackage{array}
\usepackage{amsmath}
\usepackage{amssymb}
\usepackage{setspace}
\usepackage{graphicx}
\usepackage[tight,footnotesize]{subfigure}
\usepackage{makecell,multirow,diagbox}
\usepackage[justification=centering]{caption}

\begin{document}

\title{MDBV: Monitoring Data Batch Verification for Survivability of Internet of Vehicles}

\author{Jingwei~Liu,~\IEEEmembership{Member,~IEEE,} Qingqing Li, Huijuan Cao, Rong~Sun,~\IEEEmembership{Member,~IEEE,}
        \\ Xiaojiang~Du,~\IEEEmembership{Senior Member,~IEEE}, Mohsen Guizani,~\IEEEmembership{Fellow,~IEEE}
\thanks{This work is supported in part by the Key Program of NSFC-Tongyong Union Foundation under Grant U1636209, Joint Fund of Ministry of Education of China (No.6141A02022338), the 111 Project (B08038) and Collaborative Innovation Center of Information Sensing and Understanding at Xidian University.}
\thanks{Jingwei Liu, Qingqing Li, Huijuan Cao, and Rong Sun are with the School of Telecommunications Engineering, Xidian University, Xi'an 710071, China. (e-mail: jwliu@mail.xidian.edu.cn; 15229259171@163.com; caohuijuan345@163.com; rsun@mail.xidian.edu.cn)}
\thanks{Xiaojiang Du is with the Department of Computer and Information Sciences, Temple University, Philadelphia, PA 19122, USA. (e-mail: dxj@ieee.org)}
\thanks{Mohsen Guizani is with the Department of Electrical and Computer Engineering, University of Idaho, Moscow, ID 83844, USA. (e-mail: mguizani@ieee.org)}
}

\markboth{Journal of \LaTeX\ Class Files,~Vol.~14, No.~8, June~2017}%
{Shell \MakeLowercase{\textit{et al.}}: MDBV: Monitoring Data Batch Verification for Survivability of Internet of Vehicles}
\maketitle

\begin{abstract}
 Along with the development of vehicular sensors and wireless communication technology, Internet of Vehicles (IoV) is emerging that can improve traffic efficiency and provide a comfortable driving environment. However, there is still a challenge how to ensure the survivability of IoV. Fortunately, this goal can be achieved by quickly verifying real-time monitoring data to avoid network failure. Aggregate signature is an efficient approach to realize quick data verification quickly. In this paper, we propose a monitoring data batch verification scheme based on an improved certificateless aggregate signature for IoV, named MDBV. The size of aggregated verification message is remain roughly constant even as the increasing number of vehicles in MDBV. Additionally, MDBV is proved to be secure in the random oracle model assuming the intractability of the computational Diffie-Hellman problem. In consideration of the network survivability and performance, the proposed MDBV can decrease the computation overhead and is more suitable for IoV.
\end{abstract}

\begin{IEEEkeywords}
Internet of Vehicles, Survivability, Batch Verification, Aggregate Signature, Big data.
\end{IEEEkeywords}

\IEEEpeerreviewmaketitle

\section{Introduction}

\IEEEPARstart{T}{oday}, the development of wireless networks (such as WiMAX, ad hoc, and sensor networks) has attracted worldwide attention by providing more convenient communication services. As an emerging paradigm, Internet of Vehicles (IoV) \cite{gerla2014internet} is evolving from Vehicular Ad hoc Networks (VANETs). It merges vehicles, infrastructure, human and networks to an intelligent unit that is more efficient compared with VANETs. Moreover, IoV adopts different kinds of technologies (e.g. self-organization, deep learning and cloud computing) to improve the network survivability and reliability.

\begin{figure}[tb]
  \centering
  \includegraphics[width=8.5cm]{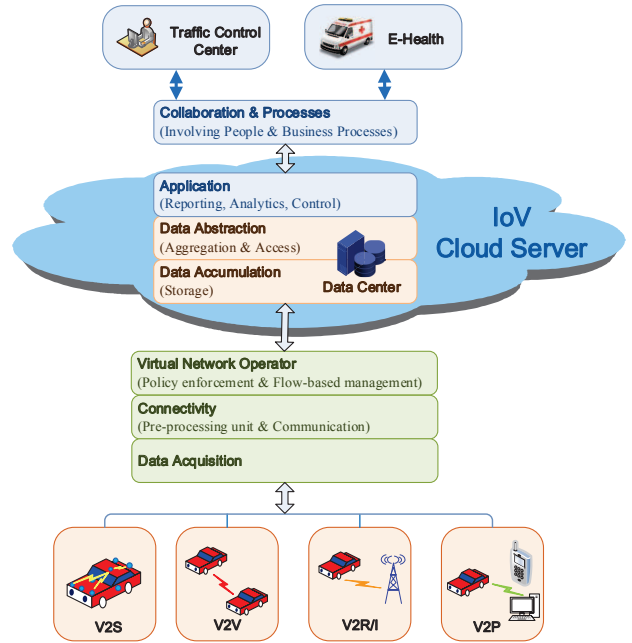}\\
  \caption{Seven-level reference model}
  \label{model}
\end{figure}

In recent years, lots of works \cite{Ahmed2016Internet, mosenia2017comprehensive, Mehmood2017Internet, yao2013lightweight, Yaqoob2017Internet, gubbi2013internet, Du2008security, Liu2018anonymous, CISCO14, du2009transactions, hei2010defending, hei2011biometric, du2008secure} have been proposed to the reliability, flexibility, survivability, and security of wireless networks. To better understand IoV, many researchers have raised several reference models: the three-level model \cite{liu2011internet}, the four-level model \cite{bonomi2013smart}, and the five-level model \cite{kaiwartya2016internet}. Combining the merits of these models, we propose a comprehensive model for IoV, as shown in Fig. \ref{model}.

Nowadays, IoV is playing an important role in smart city. It has made the traffic more efficient \cite{Lin2017Internet}. Nevertheless, for the intrinsic features of IoV, it has to face lots of challenges about the network survivability. Security and performance are two key factors of affecting the network survivability. When an malicious attacker uploads wrong information such as fake accident message, it may cause traffic control center to make incorrect decisions. So it is important to verify all of the uploaded data to enhance the survivability of IoV. However, the burden of collecting the monitoring data from different sources (such as vehicle¡¯s sensors, infrastructures, smart terminals, and so on) may increase with vehicles, and affect the transmission efficiency. In order to enhance the network survivability and prevent the data from being falsified, the protection and verification mechanisms should be considered \cite{du2007effective, xiao2007survey}. Aggregate Signature (AS) is a suitable cryptographic primitive to batch verify big data in IoV scenarios, because it can merge \emph{n} signatures on \emph{n} different messages into a single short signature. Thereby, it is deployed in MDBV to improve the survivability and performance of IoV.

In 2003, Boneh et al. proposed the first AS scheme \cite{boneh2003aggregate}. Following the original work, many aggregate schemes over different public key cryptosystems have been proposed. These schemes are widely used in vehicular communications, mobile networks, and other resource-constrained scenarios \cite{Liu2017VDAS}.

The main contribution of our paper is summarized as follows:
\begin{itemize}
\item A CLAS-based monitoring data batch verification scheme with lower computation and communication overhead is proposed to enhance the survivability of IoV scenarios.
\item A state information is introduced for vehicles to join or leave the system dynamically. Once knowing the state information, vehicles can generate their data authentication information independently.
\item The security properties of MDBV are proved in detail, assuming the hardness of the CDH problem, even though the super adversary launches the adaptive-chosen-message attack and adaptive-chosen-identity attack.
\item The performance evaluation indicates that MDBV achieves the less overhead in the phase of individual monitoring data signing and batch verification among these selected schemes, which is more compatible and preferred by vehicles and the data center respectively.
\end{itemize}

The remainder of this paper is organized as follows. Section II introduces the related work about aggregate signature. Some preliminaries are briefly introduced, involving bilinear pairing and security model in section III. In section IV, the proposed scheme MDBV is described in detail. Then, we analyze the security of MDBV in section V. The next section evaluates the performance of MDBV. Conclusion of this paper is drawn in section VII.

\section{Related Work}
Cheon et al. \cite{cheon2004new} proposed the first ID-based AS scheme in 2004. In 2007, Gong et al. \cite{gong2007two} introduced a certificateless aggregate signature to solve the key escrow problem of ID-PKC. They constructed two specific schemes using bilinear mapping that can resist to two types of attackers. Nevertheless, the size of the aggregated signature was related to user number. Since the merits of CL-PKC, soon afterwards, many CL-AS schemes \cite{liu2014secure, zhang2010efficient, xiong2013efficient, chen2015secure, du2013efficient, chen2015certificateless, kang2016secure, he2012note} were proposed. In \cite{liu2014secure}, the authors designed an efficient CL-AS scheme that required short group elements in aggregation phase and constant pairing operations in batch verification phase, whereas it cannot achieve unforgeability. Though the size of the aggregated signature was independent of the number of signers, the scheme in \cite{zhang2010efficient} required the participants to negotiate a new status information to generate an individual signature every time. In \cite{xiong2013efficient}, the scheme was more efficient than the schemes in \cite{gong2007two, zhang2010efficient}. Unfortunately, it was proved to be insecure in the case of Type II adversary in \cite{he2012note}. To address the above issues, we propose an improved CL-AS for our batch verification scheme.

\section{Preliminaries}
To better understand our scheme, we introduce some preliminaries about the properties of bilinear pairings and the security model of MDBV.

\subsection{Bilinear Pairings}
\noindent \textbf {Definition 1.} \emph{Bilinear Pairings}: $G_{1}$ and $G_{2}$ are two groups with same prime order $q$. We utilizes a bilinear pairing $e \colon G_{1}\times G_{1}\to G_{2}$ to indicate the relation of two groups, and it has the following properties:

\begin{itemize}
\item Bilinearity: We have $e(aP, bQ) = e(P, Q)^{ab}$, where $P$, $Q \in G_{1}$, random number $a$, $b\in {\rm Z}_q^\ast$;

\item Non-degeneracy: There exists $P$, $Q \in G_{1}$, such that $e(P, Q) \neq 1$;

\item Computability: There is an efficient algorithm to compute $e(P, Q)$ for any $P$, $Q \in G_{1}$.
\end{itemize}

\noindent \textbf{Definition 2.} \emph{Computational Diffie-Hellman (CDH) Problem}: Given $P, aP, bP\in G_{1}$ for any $a$, $b\in {\rm Z}_q^\ast$, output ${abP}$.

\begin{figure}[tb]
\begin{center}
\includegraphics[width=2.5in]{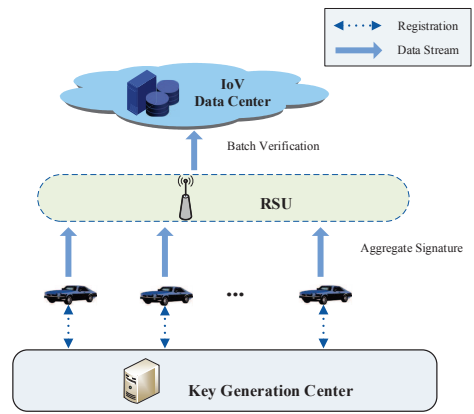}\\
\caption{The network architecture of MDBV}\label{MDBV}
\end{center}
\end{figure}

\subsection{Security Model of MDBV}
 According to the definition in \cite{huang2011certificateless}, adversaries are divided into three types: normal, strong, and super adversaries based on the different abilities of accessing signature oracles. The normal attacker is able to replace the public key of target user without obtaining his/her signature. The strong adversary can not obtain the target signer's signature unless a challenger knows the secret value associated with a replaced public key. The super adversary can also obtain the target user's signature even if the challenger doesn't know the secret value. Obviously, the super adversary has the strongest attack power. Thus, we assume there are two types of super adversaries in the proposed scheme: $\mathcal{A}_{I}$ and $\mathcal{A}_{II}$. $\mathcal{A}_{I}$ is able to replace any user's public key and get valid message signature pairs, while $\mathcal{A}_{II}$ can replace any user's public key except the target user. Here, we utilize the security model in \cite{huang2011certificateless} to prove the security of MDBV. Our scheme is proved to be secure against $\mathcal{A}_{I}$ and $\mathcal{A}_{II}$.

\section{An Efficient Monitoring Data Batch Verification for Survivability of Internet of Vehicles}
 Vehicles can gather a large number of monitoring data from their own sensors, other vehicles, infrastructure and so on, then they uploads the data to IoV data center. Therefore, when the number of vehicles is increasing, it has been a key challenge for IoV to make sure the timeliness and validity of the massive sensing data. So, we design an efficient monitoring data batch verification scheme for the survivability of IoV. To facilitate understanding, the relevant notations are listed in Table \ref{notation}.

\subsection{Design Objectives}
With the increasing of vehicles, the burden of authentication becomes heavier. To address this problem, we propose an efficient method for verifying masses of data in IoV. It can provide batch verification of monitoring data based on an improved certificateless aggregate scheme. The network architecture of MDBV is shown in Fig. \ref{MDBV}, which contains Key Generation Center (KGC), vehicles, road-side units and IoV data center. The KGC is responsible for generating the cryptographic keys of all entities in IoV.

\begin{table}
  \centering
  \caption{Notations} \label{notation}
  \footnotesize
  \tabcolsep 0.05in
  \setlength{\extrarowheight}{0.25cm}
 \begin{tabular}{c|c c}
    \hline
    \quad \quad \raisebox{0.1cm}{Notations} \quad \quad  & \quad \quad  \raisebox{0.1cm}{Description} \quad \quad \\
    \hline
      KGC                                   &   A key generation center\\
      $n$                                   &   The number of vehicles\\
    $ID_i$                                  &   The identity of vehicle $i$\\
    $\Delta$                                &   The state information\\
    $P$                                     &   A generator in cyclic group\\
    $q$                                     &   The order of cyclic group\\
    $(G_{1}, +)$                            &   A cyclic additive group\\
    $(G_{2}, \cdot)$                        &   A cyclic multiplicative group\\
    $e$                                     &   A bilinear map $e: G_{1} \times G_{1} \to G_{2}$\\
    $\textsf{H}_{1}(\cdot)$                          &   A Map-To-Point hash function\\
    $\textsf{h}_{2}(\cdot)$                          &   A secure hash function\\
    $(P_{0}, Param)$                        &   The KGC's public key and system parameters\\
    $s$                                     &   The KGC's master secret key\\
    $(P_{i}, \langle x_{i}, D_{i} \rangle)$ &   The public key and private key of vehicle $i$\\
    $\langle R, V \rangle$                  &   The aggregated signature\\
    \hline
 \end{tabular}
\end{table}

\subsection{Monitoring Data Batch Verification Scheme}
Our scheme mainly consists of five algorithms: \emph{System Setup}, \emph{Registration}, \emph{Individual Monitoring Data Signing}, \emph{Aggregation}, and \emph{Batch Verification}. A state information $\Delta$  with random length is introduced to improve security. When a vehicle enters a new area, it chooses the appropriate $\Delta$ and broadcasts it. The following is the detailed steps:

\begin{itemize}
\item[1)]{\emph{System Setup}}: KGC performs the following steps to initialize the system:
 \begin{itemize}
 \item Given the security parameter $l$, KGC creates a cyclic additive group $G_{1}$ and a cyclic multiplicative group $G_{2}$ with prime order $q (q > 2^{l})$, and generates a bilinear map $e: G_{1} \times G_{1} \to G_{2}$.
 \item KGC selects a random $s \in {\rm Z}_q^\ast$ as the system master key, and keeps $s$ in secret. Next, it computes its public key $P_{0} = sP$.
 \item KGC chooses two hash functions $\textsf{H}_{1}: \{0,1\}^* \to G_{1}$, $\textsf{h}_{2}: \{0, 1\}^* \to {\rm Z}_q^*$. Finally, KGC publishes the system parameters $Param = \{G_{1}, G_{2}, e, q, P, P_{0}, \textsf{H}_{1}, \textsf{h}_{2}\}$.
 \end{itemize}

\item[2)]{\emph{Registration}}: Upon receiving a registration request from a vehicle $i$, KGC calculates $Q_{i} = \textsf{H}_{1}(ID_{i})$ and sets $D_{i} = sQ_{i}$ the partial private key for the vehicle. Then, the vehicle picks randomly a number $x_{i} \in {\rm Z}_q^\ast$, and calculates $P_{i} = x_{i}P$, then sets $x_{i}$ and $P_{i}$ as its secret value and public key, respectively. $\langle x_{i}, D_{i} \rangle$ is its whole private key. Finally, KGC returns $\langle ID_i, Q_i, P_i \rangle$ to IoV data center via a secure channel.

\item[3)]{\emph{Individual Monitoring Data Signing}}: Before uploading the collected monitoring data to RSU, vehicles can pre-process the data and filter irrelevant information to reduce the network traffic. Then, based on the common state information $\Delta$, vehicle $i$ with identity $ID_{i}$ signs a requested monitoring $data_{i}$ using its private key $\langle x_{i}, D_{i} \rangle$ as follows:
    \begin{itemize}
        \item Choose a number $r_{i} \in {\rm Z}_q^\ast$ at random and calculate $R_{i} = r_{i}P$, $h_{i} = \textsf{h}_{2}(data_{i} \parallel \Delta \parallel ID_{i})$, $g_{i} = \textsf{h}_{2}(data_{i} \parallel \Delta \parallel P_{i})$;
        \item Compute $V_{i} = g_{i}D_{i} + (x_{i}h_{i} + r_{i})U$. Here, $U = \textsf{H}_{1}(\Delta \parallel P_{0})$;
        \item Send $\langle data_{i}, \sigma_{i}\rangle$ to RSU. Here, $\sigma_{i} = (R_{i}, V_{i})$ is the individual verification information on $data_{i}$.
    \end{itemize}

\item[4)]{\emph{Aggregation}}: Upon receiving a large number of data with individual verification information, RSU converges a collection of messages with the same state information $\Delta$. For $n$ vehicles with identities $L_{ID}=\{ID_{1}, ID_{2},\ldots,ID_{n}\}$, the public keys are $L_{PK} = \{P_{1}, P_{2}, \ldots, P_{n}\}$ and the corresponding data-signature pairs are $\langle data_{1}, \sigma_{1} \rangle, \langle data_{2}, \sigma_{2} \rangle, \ldots, \langle data_{n}, \sigma_{n} \rangle$, respectively. RSU computes $R = R_{1} + R_{2} + \ldots + R_{n}$, $V = V_{1} + V_{2} + \ldots + V_{n}$ and sets $\sigma = \langle R, V \rangle$ as the aggregated signature. Eventually, RSU forwards all data with the single verifiable signature to the data center.

\item[5)]{\emph{Batch Verification}}: To check the validity of the final signature $\sigma$ on uploaded monitoring data, the IoV data center does as follows:
\begin{itemize}
    \item Calculate $U = \textsf{H}_{1}(\Delta \parallel P_{0})$;
    \item Calculate $Q_{i} = \textsf{H}_{1}(ID_{i})$, $h_{i} = \textsf{h}_{2}(data_{i} \parallel \Delta \parallel ID_{i})$, $g_{i} = \textsf{h}_{2}(data_{i} \parallel \Delta \parallel P_{i})$ for all $1\leqslant i\leqslant n$,;
    \item Verify the equation $e(V, P) = e(\sum_{i = 1}^{n} g_{i}Q_{i}, P_{0}) e(\sum_{i = 1}^{n} h_{i}P_{i} + R, U)$. If it holds, the uploaded data is valid. Otherwise, the data center refuses to accept these data.
 \end{itemize}
\end{itemize}

\begin{figure*}[tb]
\begin{center}
\includegraphics[width=16cm]{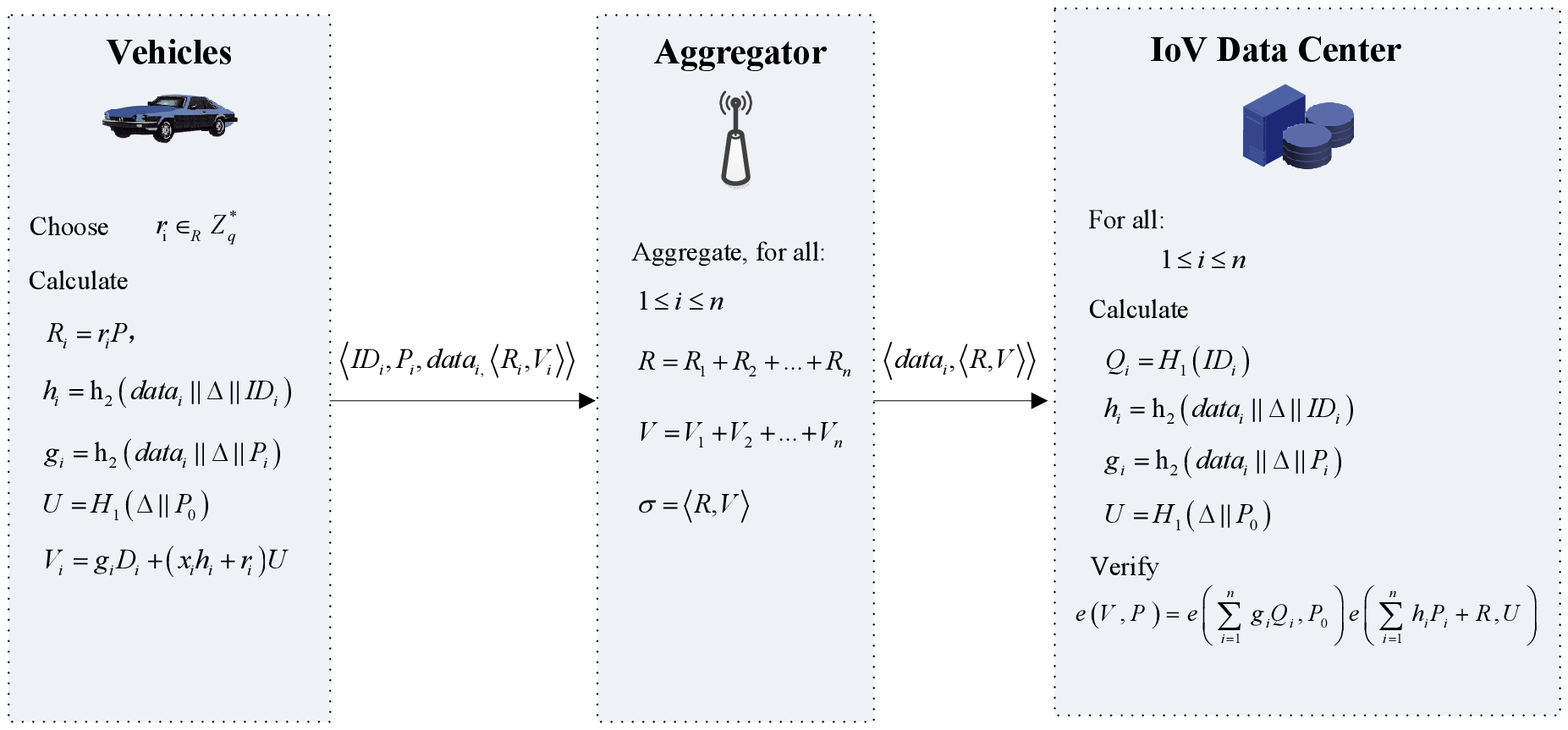}\\
\caption{The flowchart of MDBV}\label{Aggregate}
\end{center}
\end{figure*}

Fig. \ref{Aggregate} indicates the flowchart of MDBV.

\subsection{Correctness}
The correctness of MDBV is proved as follows:

\noindent \emph{For individual verification}: We can verify the individual data via the following equation:
\begin{eqnarray}
\nonumber  e(V_{i},P)&=& e(g_{i}D_{i}+(x_{i}h_{i}+r_{i})U,P) \\
\nonumber            &=& e(g_{i}Q_{i},P_{0})e((x_{i}h_{i}+r_{i})P,U) \\
\nonumber            &=& e(g_{i}Q_{i},P_{0})e(x_{i}h_{i}P+r_{i}P,U) \\
\nonumber            &=& e(g_{i}Q_{i},P_{0})e(h_{i}P_{i}+R_{i},U)
\end{eqnarray}

\noindent \emph{For batch verification}: From above, we have the following equation:
\begin{eqnarray}
\nonumber e(V,P) &=& e(\sum_{i = 1}^{n}V_{i},P) \\
\nonumber        &=& e(\sum_{i = 1}^{n}(g_{i}D_{i}+(x_{i}h_{i}+r_{i})U),P)\\
\nonumber        &=& e(\sum_{i = 1}^{n}g_{i}D_{i},P)e(\sum_{i = 1}^{n}(x_{i}h_{i}+r_{i})U,P)\\
\nonumber        &=& e(\sum_{i = 1}^{n}g_{i}Q_{i},P_{0})e(\sum_{i = 1}^{n}(h_{i}P_{i}+R_{i}),U)\\
\nonumber        &=& e(\sum_{i = 1}^{n}g_{i}Q_{i},P_{0})e(\sum_{i = 1}^{n}h_{i}P_{i}+R,U)
\end{eqnarray}

\section{Security Analysis}
The security analysis of MDBV is given in this section, including security proof and security property.

\subsection{Security Proof}
To prove the existential unforgeability of MDBV, we introduce two types of super adversaries $\mathcal{A}_{I}$ and $\mathcal{A}_{II}$ to play the games with a challenger respectively.

\noindent \textbf{Theorem 1.}  \emph {MDBV is existential unforgeable against the adversaries $\mathcal{A}_I$ in ROM assuming the hardness of CDHP.}

\noindent \textbf{Proof:} $\mathcal{C}$ is a challenger try to solve a random CDH instance $(P,aP,bP)$. Adversary $\mathcal{A}_{I}$ can help $\mathcal{C}$ achieve this goal in the following games.

\emph{Setup}: $\mathcal{C}$ sets $P_{0}=aP$ and $Param = \{G_{1}, G_{2}, e, q, P, P_{0}, \textsf{H}_{1}, \textsf{h}_{2}\}$, and returns $Param$ to $\mathcal{A}_{I}$.

\emph{Simulation}: $\textsf{H}_{1}$ and $\textsf{h}_{2}$ are two hash functions. Meanwhile, $\mathcal{C}$ keeps two lists $L_{\textsf{H}_{1}}$ and $L_{\textsf{h}_{2}}$. $\mathcal{A}_{I}$ can adaptively perform the following queries.

\begin{itemize}
\item[1)]\emph{Register-query}: $\mathcal{C}$ first randomly chooses $j \in \{1, 2,\ldots, n\}$, then $\mathcal{A}_{I}$ makes \emph{Register-query} on an identity $ID_i$. If $i \neq j$, $\mathcal{C}$ picks $x_{i}, y_{i} \in {\rm Z}_q^\ast$ at random, where the tuple $\langle ID_{i}, Q_{i}, P_{i}, D_{i}, x_{i}, y_{i} \rangle$ is not in $L_R$, and computes $Q_{i} = y_{i}P$, $P_{i} = x_{i}P$, $D_{i} = y_{i}P_{0}$; Otherwise, $\mathcal{C}$ randomly chooses $Q_{j} = bP \in G_1$, $x_{j} \in {\rm Z}_q^\ast$, and sets $D_{j} = ``unknown"$, $P_{j} = x_{j}P$. Then, $\mathcal{C}$ inserts $\langle ID_{i}, Q_{i}, P_{i}, D_{i}, x_{i}, y_{i} \rangle$ to $L_R$ and returns $Q_{i}$ and $P_{i}$ as the answer.

\item[2)]\emph{Partial-Private-Key-query}: $\mathcal{A}_{I}$ makes a query on an identity $ID_i$. When $i \neq j$, $\mathcal{C}$ scans $L_R$, and responds $D_{i}$ to $\mathcal{A}_{I}$. Otherwise, $\mathcal{C}$ aborts.

\item[3)]\emph{Public-Key-Replace-query}: $\mathcal{A}_{I}$ asks a question on $\langle ID_{i},P_{i}^\prime \rangle$, $\mathcal{C}$ scans $L_R$, replacing $\langle P_{i}, x_{i}, y_{i} \rangle$ with $\langle P_{i}^\prime, ``unknown", y_{i} \rangle$.

\item[4)]\emph{Secret-Value-query}: $\mathcal{A}_{I}$ asks a question on an identity $ID_i$. $\mathcal{C}$ first searches $L_R$, if $x_{i} \neq ``unknown"$, it sends $x_{i}$ to $\mathcal{A}_{I}$; Otherwise, it sends $``unknown"$.

\item[5)]$\textsf{H}_{1}$-\emph{query}: $\mathcal{A}_{I}$ can make a query on $\langle \Delta_i, P_{0} \rangle$. If the tuple $\langle \Delta_i, U_{i}, \lambda_{i} \rangle$ is not in $L_{\textsf{H}_{1}}$, $\mathcal{C}$ chooses $\lambda_{i} \in {\rm Z}_q^\ast$ . Then, it sets $U_{i} = \lambda_{i}P - P_{0}$, and sends it to $\mathcal{A}_{I}$. Finally, $\langle \Delta_i, U_{i}, \lambda_{i} \rangle$ will be added to $L_{\textsf{H}_{1}}$.

\item[6)]$\textsf{h}_{2}$-\emph{query}: $\mathcal{A}_{I}$ asks a question on any $(0, 1)^*$. If $L_{\textsf{h}_{2}}$ contains tuples $\langle data_{i}, \Delta_i, ID_{i}, h_{i} \rangle$ or $\langle data_{i}, \Delta_i, P_{i}, g_{i} \rangle$, $\mathcal{C}$ chooses $h_{i}$ or $g_{i} \in {\rm Z}_q^\ast$. Then, it returns $h_{i}$ or $g_{i}$ to $\mathcal{A}_{I}$ and inserts the tuple to $L_{\textsf{h}_{2}}$.

\item[7)]\emph{Individual-Signing-query}: $\mathcal{A}_{I}$ makes a signing query on any $\langle data_{i}, \Delta_i, ID_{i}, P_{i} \rangle$, $\mathcal{C}$ does as follows:
\begin{enumerate}
  \item [(a)] Select $r_{i},h_{i},g_{i}\in {\rm Z}_q^\ast$ randomly, while $h_{i}$ and $g_{i}$ are not in $L_{\textsf{h}_{2}}$.
  \item [(b)] Compute $R_{i}=r_{i}P+g_{i}Q_{i}-h_{i}P_{i}$.
  \item [(c)] Search $U_{i} = \lambda_{i}P - P_{0}$ in $L_{\textsf{H}_{1}}$ and compute $V_{i} = \lambda_{i}g_{i}Q_{i} + r_{i}\lambda_{i}P - r_{i}P_{0}$.
\end{enumerate}
$\mathcal{C}$ stores the above information in the relevant lists and sends $\langle R_{i}, V_{i} \rangle$ to $\mathcal{A}_{I}$. $\langle R_{i}$, $V_{i}\rangle$ is valid, since
\begin{eqnarray}
\nonumber & & e(g_{i}Q_{i},P_{0})e(h_{i}P_{i}+R_{i},U_i)\\
\nonumber &=& e(g_{i}Q_{i},P_{0})e(r_{i}P+g_{i}Q_{i},\lambda_{i}P - P_{0})\\
\nonumber &=& e(\lambda_{i}r_{i}P+\lambda_{i}g_{i}Q_{i},P)e(r_{i}P, -P_{0})\\
\nonumber &=& e(\lambda_{i}g_{i}Q_{i} + r_{i}\lambda_{i}P-r_{i}P_{0}, P)\\
\nonumber &=& e(V_{i},P)
\end{eqnarray}

\item[8)]\emph{Forgery}: Finally, $\mathcal{A}_{I}$ returns $n$ vehicles, whose identities are $L_{ID} = \{ID_{1}, ID_{2}, \ldots, ID_{n}\}$ and corresponding public keys are $L_{PK} = \{P_{1}, P_{2}, \ldots, P_{n}\}$, $n$ monitoring data $\langle data_{1}, data_{2}, \ldots, data_{n} \rangle$, a same state information $\Delta$, and a forged aggregate signature $\sigma$. Moreover, the aggregate signature must satisfy the following conditions:
\begin{enumerate}
  \item [(a)] $e(V,P)=e(\sum_{i = 1}^{n}g_{i} Q_{i},P_{0})e(\sum_{i = 1}^{n}h_{i} P_{i} + R, U)$.
  \item [(b)] There exists at least an identity $ID_{k} \in L_{ID}$ has not made \emph{Partial-Private-Key-query} and \emph{Individual-Signing-query} on $\langle data_{k}, \Delta, ID_{k}, P_{k} \rangle$.
\end{enumerate}

According to Forking lemma \cite{pointcheval2000security}, $\mathcal{A}_{I}$ can forge a new signature $\sigma^\prime = \langle R, V^\prime \rangle$ through replaying technology with the same random tape but a different response to $\textsf{h}_{2}$. In this process, if $i\in\{1,2,\ldots,n\}\backslash\{k\}$, $g_{i} = g_{i}^\prime$; otherwise, $g_{k} \neq g_{k}^\prime$. Hence, we have the following equations:
\begin{equation}
 \begin{split}\nonumber
 \begin{cases}
    e(V, P) = e(\sum_{i = 1}^{n}g_{i} Q_{i},P_{0})e(\sum_{i = 1}^{n}h_{i} P_{i}+R, U)\\
    e(V^\prime, P) = e(\sum_{i=1}^{n}g_{i}^\prime Q_{i},P_{0})e(\sum_{i = 1}^{n}h_{i} P_{i}+R, U)
 \end{cases}
 \end{split}
\end{equation}

If $ID_{k} = ID_{j}$, meaning $Q_{k} = Q_{j} = bP$, $\mathcal{C}$ calculates $abP = (g_{j} - g_{j}^\prime)^{-1}(V - V^\prime)$ and sets $abP$ as the solution of CDHP; otherwise, it aborts.

\end{itemize}

Through the analysis on the attack of adversary $\mathcal{A}_{I}$, we can also deduce Theorem 2 in the similar way.

\noindent \textbf{Theorem 2.} \emph {MDBV is existentially unforgeable against the adversaries $\mathcal{A}_{II}$ in ROM assuming the intractability of CDHP.}

\noindent \textbf{Proof:} $\mathcal{C}$ plays the same role as in Theorem 1. Adversary $\mathcal{A}_{II}$ can help $\mathcal{C}$ to solve the CDH problem.

\emph{Setup}: $\mathcal{C}$ randomly selects a number $s \in {\rm Z}_q^\ast$ as the system secret key and calculates its public key $P_{0}=sP$. Then, it sends $Param = \{G_{1}, G_{2}, e, q, P, P_{0}, \textsf{H}_{1}, \textsf{h}_{2}\}\cup \{s\}$ to $ \mathcal{A}_{II}$.

\emph{Simulation}: $\mathcal{A}_{II}$ can adaptively question all the following queries except Partial-Private-Key-query. $\mathcal{C}$ holds two lists $L_{\textsf{H}_{1}}$ and $L_{\textsf{h}_{2}}$.

\begin{itemize}
\item[1)]\emph{Register-query}: $\mathcal{C}$ first randomly picks $j \in \{1, 2,\ldots, n\}$, then $\mathcal{A}_{II}$ can make a \emph{Register-query} on an identity $ID_i$. $\mathcal{C}$ picks $x_{i}, y_{i} \in {\rm Z}_q^\ast$ at random, where $y_{i}$ does not exist in $L_R$. If $i \neq j$, $\mathcal{C}$ calculates $Q_{i} = y_{i}P$, $P_{i} = x_{i}P$; otherwise, $\mathcal{C}$ computes $Q_{j} = y_{j}P$, and sets $P_{j} = x_{j}bP$. Then, $\mathcal{C}$ inserts the tuple $\langle ID_{i}, Q_{i}, P_{i}, D_{i}, x_{i}, y_{i} \rangle$ to $L_R$ and returns $Q_{i}$ and $P_{i}$ as answer.
\item[2)]\emph{Public-Key-Replace-query}: $\mathcal{A}_{II}$ makes a query on $\langle ID_{i},P_{i}^\prime \rangle$, then $\mathcal{C}$ checks it in $L_R$ and does as follows:
\begin{enumerate}
  \item [(a)] If $i\neq j$, $\mathcal{C}$ performs a public key replacement. Namely, $\mathcal{C}$ replaces $\langle P_{i}, x_{i}, y_{i}\rangle$ with $\langle P_{i}^\prime, ``unknown", y_{i}\rangle$;
  \item [(b)] If $ i = j $, $\mathcal{C}$ aborts.
\end{enumerate}
\item[3)]\emph{Secret-Value-query}: $\mathcal{A}_{II}$ can query on any $ID_{i}$. Then, $\mathcal{C}$ checks it in $L_R$ and does as follows:
\begin{enumerate}
  \item [(a)] If $i \neq j$ and $x_{i} \neq ``unknown"$, $\mathcal{C}$ returns $x_{i}$ to $\mathcal{A}_{II}$. Otherwise, it outputs $``unknown"$;

  \item [(b)] If $ i = j $, $\mathcal{C}$ aborts.
\end{enumerate}
\item[4)]$\textsf{H}_{1}$-\emph{query}: $\mathcal{A}_{II}$ can ask a question on $\langle \Delta_i, P_{0} \rangle$. $\mathcal{C}$ chooses $\lambda_{i} \in {\rm Z}_q^\ast$. Next, it computes $U_{i} = \lambda_{i}P - aP$, and sends it to $\mathcal{A}_{II}$. Finally, $\langle \Delta_i, U_{i}, \lambda_{i} \rangle$ will be inserted to $L_{\textsf{H}_{1}}$.
\item[5)]$\textsf{h}_{2}$-\emph{query}: $\mathcal{A}_{II}$ queries on an identity $ID_{i}$, $\mathcal{C}$ picks $h_{i}$ or $g_{i} \in {\rm Z}_q^\ast$. Then, it sends $h_{i}$ or $g_{i}$ to $\mathcal{A}_{II}$ and inserts $\langle data_{i}, \Delta_i, ID_{i}, h_{i} \rangle$ or $\langle data_{i}, \Delta_i, P_{i}, g_{i} \rangle$ to $L_{\textsf{h}_{2}}$.
\item[6)]\emph{Individual-Signing-query}: $\mathcal{A}_{II}$ makes a signing query on any $\langle data_{i}, \Delta_i, ID_{i}, P_{i} \rangle$, $\mathcal{C}$ does as follows:
\begin{enumerate}
  \item [(a)] Select $r_{i},h_{i},g_{i}\in {\rm Z}_q^\ast$ randomly, while $h_{i}$ and $g_{i}$ are not in $L_{\textsf{h}_{2}}$.
  \item [(b)] Compute $R_{i}=r_{i}P - h_{i}P_{i}$.
  \item [(c)] Search $U_{i} = \lambda_{i}P - aP$ in $L_{\textsf{H}_{1}}$ and compute $V_{i} = sg_{i}Q_{i} + r_{i}\lambda_{i}P - r_{i}aP$.
\end{enumerate}

$\mathcal{C}$  stores the above information to the relevant lists and sends $\langle R_{i}, V_{i} \rangle$ to $\mathcal{A}_{II}$. According to the equation $P_{0} = sP$ and the individual signing verification algorithm, we can check if $\langle R_{i}$, $V_{i}\rangle$ is valid, since
\begin{eqnarray}
\nonumber & & e(g_{i}Q_{i},P_{0})e(h_{i}P_{i}+R_{i},U_i)\\
\nonumber &=& e(g_{i}Q_{i},sP)e(h_{i}P_{i}+R_{i},\lambda_{i}P - P_{0})\\
\nonumber &=& e(sg_{i}Q_{i},P)e(r_{i}P,\lambda_{i}P-P_{0})\\
\nonumber &=& e(sg_{i}Q_{i},P)e(r_{i}(\lambda_{i}P-P_{0}),P)\\
\nonumber &=& e(sg_{i}Q_{i},P)e(r_{i}\lambda_{i}P-r_{i}P_{0},P)\\
\nonumber &=& e(sg_{i}Q_{i}+r_{i}\lambda_{i}P-r_{i}P_{0},P)\\
\nonumber &=& e(V_{i},P)
\end{eqnarray}

\item[7)]\emph{Forgery}: Finally, $\mathcal{A}_{I}$ returns $n$ vehicles, whose identities are $L_{ID} = \{ID_{1}, ID_{2}, \ldots, ID_{n}\}$ and corresponding public keys are $L_{PK} = \{P_{1}, P_{2}, \ldots, P_{n}\}$, $n$ monitoring data $\langle data_{1}, data_{2}, \ldots, data_{n} \rangle$, a same $\Delta$, and a forged $\sigma=(R,V)$ that satisfies the following cases:
\begin{enumerate}
  \item [(a)] $e(V,P)=e(\sum_{i = 1}^{n}g_{i} Q_{i},P_{0})e(\sum_{i = 1}^{n}h_{i} P_{i} + R, U)$.
  \item [(b)] There is at least an identity $ID_{k} \in L_{ID}$, which has not made \emph{Individual-Signing-query} on $\langle data_{k}, \Delta, ID_{k}, P_{k} \rangle$.
\end{enumerate}
Here, according to Forking lemma, $\mathcal{A}_{I}$ can forge a new signature $\sigma^\prime = \langle R, V^\prime \rangle$ using replaying technology. In this process, if $i\in\{1,2,\ldots,n\}\backslash\{k\}$, $h_{i} = h_{i}^\prime$; otherwise, $h_{k} \neq h_{k}^\prime$.  Hence, we have the following equations:
\begin{equation}
 \begin{split}\nonumber
 \begin{cases}
    e(V, P) = e(\sum_{i = 1}^{n}g_{i} Q_{i},P_{0})e(\sum_{i = 1}^{n}h_{i} P_{i}+R, U)\\
    e(V^\prime, P) = e(\sum_{i=1}^{n}g_{i}Q_{i},P_{0})e(\sum_{i = 1}^{n}h_{i}^\prime P_{i}+R, U)
 \end{cases}
 \end{split}
\end{equation}

Then the below equation is obtained.
\begin{equation*}
  e((h_{j}-h_{j}^\prime)P_{j},P) = e(V-V^\prime,P)
\end{equation*}
If $ID_{k} = ID_{j}$, then $U_{i} = \lambda_{i}P - aP$ and $P_{k} = P_{j} = x_{j}bP$. $\mathcal{C}$ could solve the CDHP by computing $abP = \lambda_{i}x_{j}P + \lambda_{i}bP - ax_{j}P - (V-V^\prime)(h_{j}-h_{j}^\prime)^{-1}$ according to the above equation. Otherwise, it aborts.

\end{itemize}

\begin{table*}
  \centering
  \caption{Security property comparisons between different schemes} \label{security}
  \setlength{\extrarowheight}{0.25cm}
  \footnotesize
  \tabcolsep 0.05in
  \begin{tabular}{|c|c|c|c|c|c|c|c|c|c|c|}
  \hline
  \multicolumn{2}{|c|}{Schemes} & GLHC\cite{gong2007two}-1 & GLHC\cite{gong2007two}-2 & ZQWZ\cite{zhang2010efficient} & XGCL\cite{xiong2013efficient} & CWZY\cite{chen2015secure} & DHW\cite{du2013efficient} & LYX\cite{lu2012provably} & YZW\cite{Yang2014Certificateless} & MDBV \\
  \hline
  \multirow{4}*{Adversary $\mathcal{A}_I$} & Normal & \quad & \quad & \quad & \quad & \quad & $\surd$ & $\surd$ & \quad & \quad \\
  \cline{2-11}
                                           & Strong & $\surd$ & $\surd$ & \quad & \quad & \quad & \quad & \quad & \quad & \quad \\
  \cline{2-11}
                                           & Super  & \quad & \quad & $\surd$ & $\surd$ & $\surd$ & \quad & \quad & $\surd$ & $\surd$ \\
  \cline{2-11}
                                           & Security Property & weak & weak & strong & strong & strong & weak & strong & strong & strong \\
  \hline
  \multirow{4}*{Adversary $\mathcal{A}_{II}$} & Normal & $\surd$ & $\surd$ & $\surd$ & \quad & \quad & $\surd$ & $\surd$ & $\surd$ & \quad \\
  \cline{2-11}
                                            & Strong   & \quad & \quad & \quad & \quad & \quad & \quad & \quad & \quad & \quad \\
  \cline{2-11}
                                            & Super    & \quad & \quad & \quad & $\surd$ & $\surd$ & \quad & \quad & \quad & $\surd$ \\
  \cline{2-11}
                                            & Security Property & weak & weak & weak & weak & strong & weak & weak & weak & strong \\
  \hline
  \end{tabular}
\end{table*}

\subsection{Security Property}
 We have the security property comparison of our scheme with the existing schemes \cite{gong2007two, zhang2010efficient, xiong2013efficient, chen2015secure, du2013efficient, lu2012provably, Yang2014Certificateless} in this part. As shown in Table \ref{security}, the schemes in \cite{du2013efficient, lu2012provably} can only resist the normal adversary $\mathcal{A}_I$ and $\mathcal{A}_{II}$, and their security is weak. Also, the schemes \cite{gong2007two}-1 and \cite{gong2007two}-2 have weak security property though they have the resistance to the strong adversary $\mathcal{A}_I$ and the normal adversary $\mathcal{A}_{II}$. Meanwhile, although the resistance to the first type of adversary $\mathcal{A}_I$ is stronger in \cite{zhang2010efficient, Yang2014Certificateless}, both of them can't resist the second normal adversary ($\mathcal{A}_{II}$). Moreover, the scheme in \cite{xiong2013efficient} can resist the super adversary $\mathcal{A}_I$ but it is not strong enough to resist the second super adversary ($\mathcal{A}_{II}$). From Theorem 1 and Theorem 2, MDBV can resist both two types of super adversaries $\mathcal{A}_I$ and $\mathcal{A}_{II}$, it achieves better security property compared with other schemes.

\section{Performance Evaluation}
In this section, we compare MDBV with several existing schemes \cite{zhang2010efficient, chen2015secure, du2013efficient, chen2015certificateless} in terms of performance evaluation. The details are as follows:

\subsection{Computation Overhead}
We set up a simulation environment to evaluate the computation costs. Firstly, we test the performance in terms of computation and then closely compare MDBV with its non-aggregate mode, named un-Agg mode, and other four schemes. In un-Agg mode, all data is verified by RSU one by one. Then, we analyze the computation efficiency of all schemes.

\begin{table*}
  \centering
  \caption{Comparisons among six schemes} \label{Complexity}
  \setlength{\extrarowheight}{0.25cm}
  \footnotesize
  \tabcolsep 0.05in
  \begin{tabular}{c|ccc}
  \hline
  \quad \quad \quad \raisebox{0.1cm}{Schemes} \quad \quad \quad & \quad \quad \quad \raisebox{0.1cm}{Signing} \quad \quad \quad & \quad \quad \quad \raisebox{0.1cm}{Batch verification} \quad \quad \quad & \quad \quad \quad \raisebox{0.1cm}{Length} \quad \quad \quad \\
  \hline
  ZQWZ\cite{zhang2010efficient}   &   $5M+3H$  &   $5P+2nM+(2n+3)H$        &   $2L$     \\
  CWZY\cite{chen2015secure}   &   $4M+2H$  &   $4P+2nM+(n+2)H$         &   $(n+1)L$ \\
  DHW\cite{du2013efficient}    &   $4M+2H$  &   $4P+2nM+(n+2)H$         &   $2L$     \\
  CTMHH\cite{chen2015certificateless}  &   $4M+2H$  &   $4P+2nM+2nH$            &   $2L$     \\
  un-Agg mode        &   $3M+1H$  &   $3nP+2nM+2nH$           &   $2nL$  \\
  MDBV               &   $3M+1H$  &   $3P+2nM+(n+1)H$         &   $2L$     \\
  \hline
  \end{tabular}
\end{table*}

Table \ref{Complexity} shows the computation complexity among the selected schemes. Let ``$P$" denote the bilinear pairing in $G_{1}$, ``$M$" denote multiplication in $G_{1}$, ``$H$" denote the Map-To-Point operation, and ``$L$" denote the length of the elements in $G_{1}$. From Table \ref{Complexity}, MDBV only involves three ``$M$" and one ``$H$" in the individual signing stage, and requires less computation complexity than the other schemes in the phase of batch verification. Moreover, we find that all except for the scheme in \cite{chen2015secure} have the fixed length of the batch verification signature---$2L$. We will describe the trend of computation overheads with the number of vehicles in the following part.

\subsubsection{Platform setup}
In order to measure the computation overhead of the selected schemes, we set Raspberry Pi 3B+ as the hardware environment that runs Raspbian GNU/Linux 8.0 (jessie) over Broadcom BCM2837 64 Bit ARMv7 Quad Core 1.2GHz Processor with 1GB 400MHz SDRAM. The simulation is implemented based on the GNU Multiple Precision Arithmetic (GMP) library and Pairing Based Cryptography (PBC) library. The elliptic curve is $y^2 = x^3 + x$, in which the pairing operation is symmetric and the embedding degree k is 2. We run each scheme 1000 times to eliminate the randomness of the results.

\subsubsection{Simulation results and analysis}
There are three brief kinds of cryptographic operations causing major computation overhead in these schemes: ``$M$", ``$P$", and ``$H$" operations. Table \ref{cryptographic operations} shows that the time consumption of the three basic cryptographic operations. The time consumption on individual data signing is shown in Table \ref{signing}.

\begin{table}
  \centering
  \caption{Running time of basic operations} \label{cryptographic operations}
  \tabcolsep 0.05in
  \setlength{\extrarowheight}{0.25cm}
  \footnotesize
  \begin{tabular}{c|ccc}
     \hline
     \quad \raisebox{0.1cm}{Operations} \quad & \quad \raisebox{0.1cm}{Multiplication} \quad & \quad \raisebox{0.1cm}{Map-To-Point} \quad & \quad \raisebox{0.1cm}{Pairing} \quad     \\
     \hline
     \quad \raisebox{0.1cm}{Time(ms)}   \quad & \quad \raisebox{0.1cm}{10.087}          \quad & \quad \raisebox{0.1cm}{23.417} \quad & \quad \raisebox{0.1cm}{15.063}  \quad     \\
     \hline
  \end{tabular}
\end{table}

\begin{table}
  \centering
  \caption{Time consumption on individual data signing} \label{signing}
  \footnotesize
  \tabcolsep 0.05in
  \setlength{\extrarowheight}{0.25cm}
 \begin{tabular}{c|c c}
    \hline
    \quad \quad \raisebox{0.1cm}{Schemes} \quad \quad  & \quad \quad  \raisebox{0.1cm}{Time consumption (ms)} \quad \quad \\
    \hline
    ZQWZ\cite{zhang2010efficient}  &   122.529\\
    CWZY\cite{chen2015secure}  &   88.571\\
    DHW\cite{du2013efficient}   &   87.664\\
    CTMHH\cite{chen2015certificateless} &   88.978\\
    MDBV              &   54.324\\
    \hline
 \end{tabular}
\end{table}

\begin{figure*}[tb]
\centerline{
\subfigure[Time consumption on individual data signing]{\includegraphics[width=8cm]{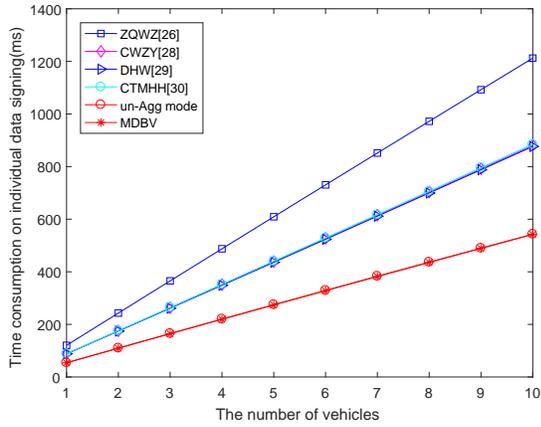}%
\label{signing_with_users}} %
\hspace{3mm}
\subfigure[Time consumption on batch verification]{\includegraphics[width=8.2cm]{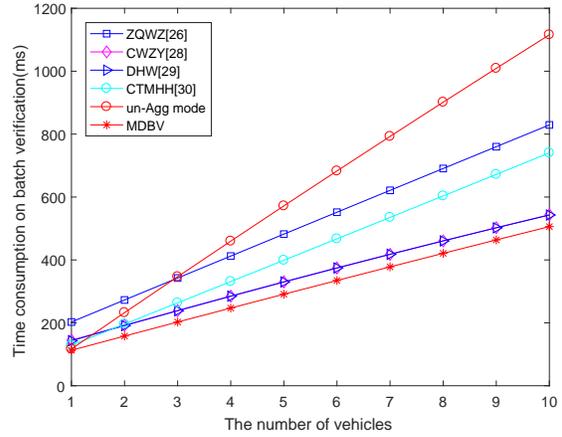}%
\label{verification_with_users}} %
}
\setlength{\belowcaptionskip}{-2 mm}
\caption{Time consumption comparisons between different schemes}
\label{comparison}
\vspace{-1mm}
\end{figure*}

We can easily calculate the time consumption of the ``\emph{Individual monitoring data signing}'' stage and the ``\emph{Batch verification}'' stage based on the running time of the three basic cryptographic operations. Fig. \ref{signing_with_users} shows that MDBV takes less computation overhead with the increasing number of vehicles than the other schemes \cite{zhang2010efficient, chen2015secure, du2013efficient, chen2015certificateless} in the individual data signing stage. And, it also requires less time overhead on batch verification than the other four selected schemes from Fig. \ref{verification_with_users}. Meanwhile, the efficiency of batch verification in MDBV is much higher than that in un-Agg mode.

\subsection{Communication Overhead}
To evaluate the communication overhead, we mainly analyze the message size in the communication process between RSU and IoV data center.

A point $P$ on an elliptic curve is represented by coordinates $(x, y)$ over a finite field $F_p$. Once a coordinate $x$ or $y$ is given, the point $P = (x, y)$ on a specific elliptic curve, such as $y^2 = x^3 + x$, can be easily constructed. Thus, when a vehicle tries to send a point $P = (x, y)$, it only needs to transmit the single coordinate $x$ or $y$ to reduce the communication overhead. Meanwhile, because the group order $q$ is 160 bits long and the order of the base field $p$ is 512 bits long over the elliptic curve, the length of point $P$ is 512 bits or 64 bytes.

In MDBV, the signature $\sigma_{i} = (R_{i}, V_{i})$ of vehicle $i$ consists of two points on the elliptic curve. We use $S$ to denote the length of $data_{i}$. And, we utilize a similar approach in \cite{ren2007broadcast, shim2013eibas, li2016secure} to calculate the message size of MDBV, assuming that $S$ is 160 bits. So, in MDBV, the total communication overhead for a signed data is 148 bytes, as $\left| R_{i} \right| + \left| V_{i} \right| + \left| data_{i} \right| = 64 + 64 + 20$ bytes. If we adopt other supersingular elliptic curves, like the scheme in \cite{boneh2001short}, the total communication overhead will be reduced to 60 bytes, as $\left| R_{i} \right| + \left| V_{i} \right| + \left| data_{i} \right| = 20 + 20 + 20$ bytes.

\begin{figure*}[tb]
\centerline{
\subfigure[Energy consumption on communication]{\includegraphics[width=8cm]{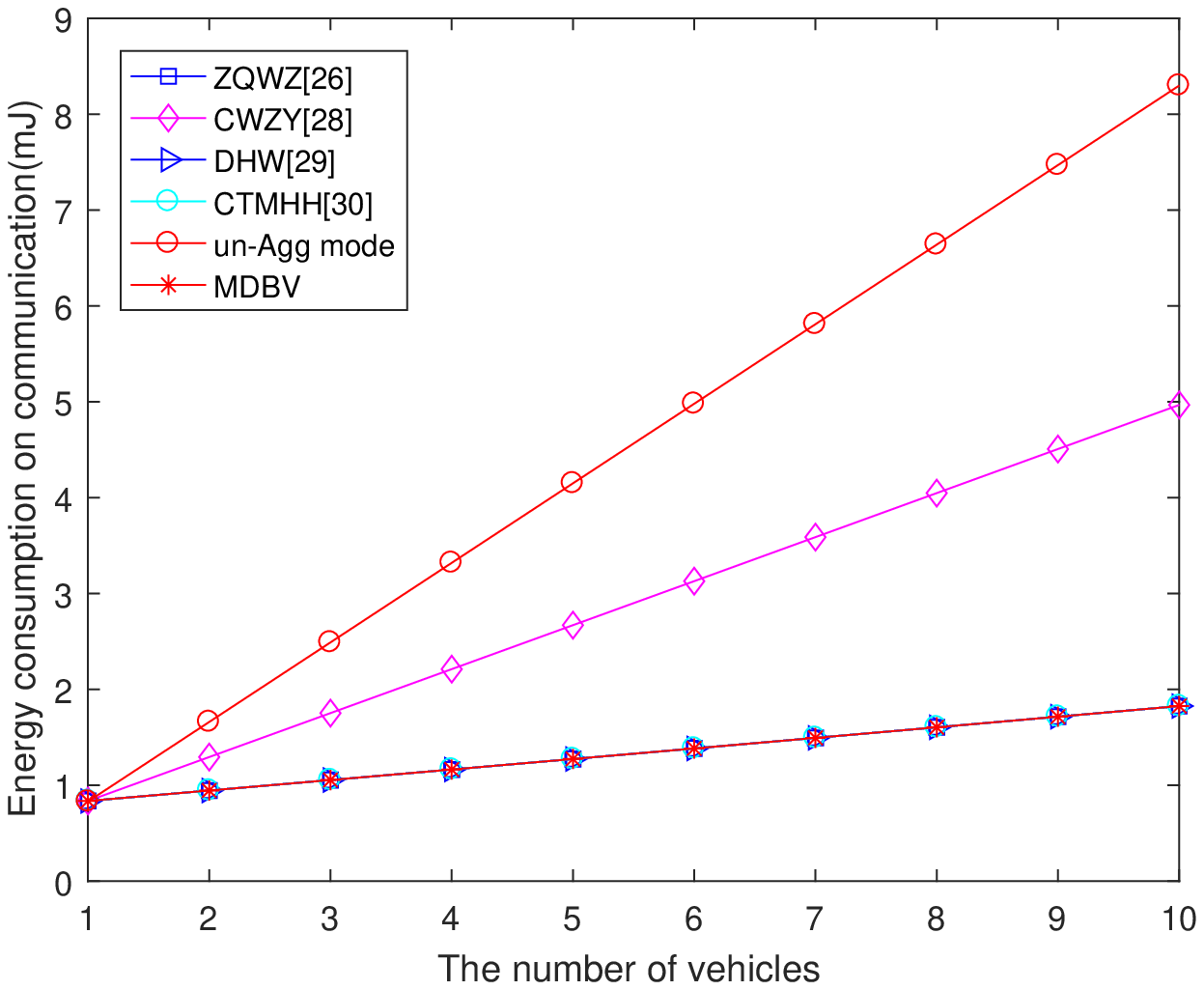}%
\label{communication}} %
\hspace{3mm}
\subfigure[Total energy consumption]{\includegraphics[width=8.2cm]{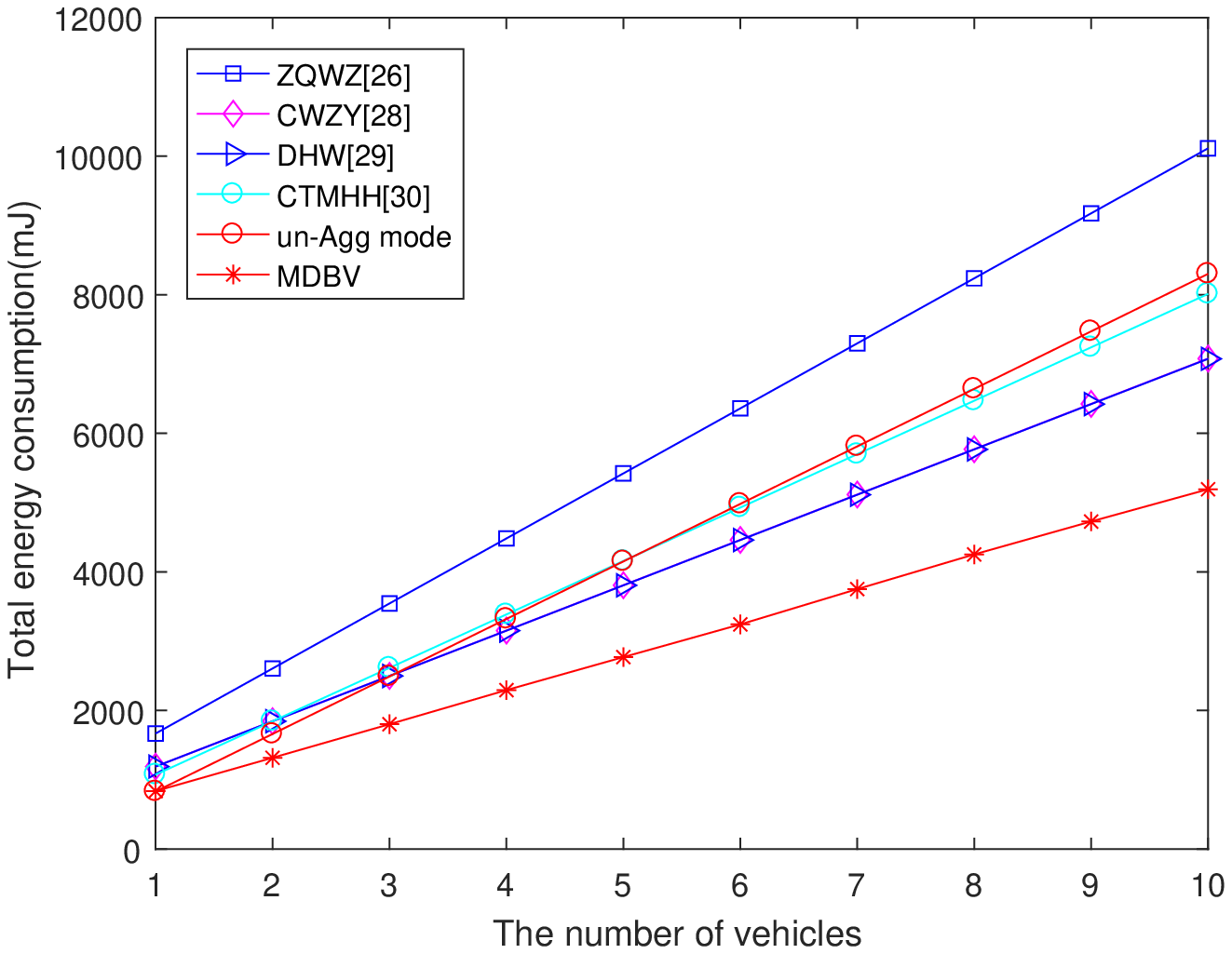}%
\label{total energy}} %
}
\setlength{\belowcaptionskip}{-2 mm}
\caption{The comparisons on energy consumption between different schemes}
\label{comparison}
\vspace{-1mm}
\end{figure*}

On account that all schemes have the same traffic between vehicles and RSU, we mainly analyze the communication overhead between RSU and IoV data center. Fig. \ref{communication} shows the comparison on communication overhead in different schemes. Among these schemes, MDBV has a fixed length of $2L$ (a constant) for data verification, as shown in Table \ref{Complexity}. Therefore, the length of authentication message in ZQWZ \cite{zhang2010efficient}, DHW \cite{du2013efficient}, CTMHH \cite{chen2015certificateless} and MDBV is $2L+n\times S$. It is far shorter than that of CWZY\cite{chen2015secure} and un-Agg mode.

\subsection{Energy Cost}
\begin{figure}[tb]
\begin{center}
\includegraphics[width=8.5cm]{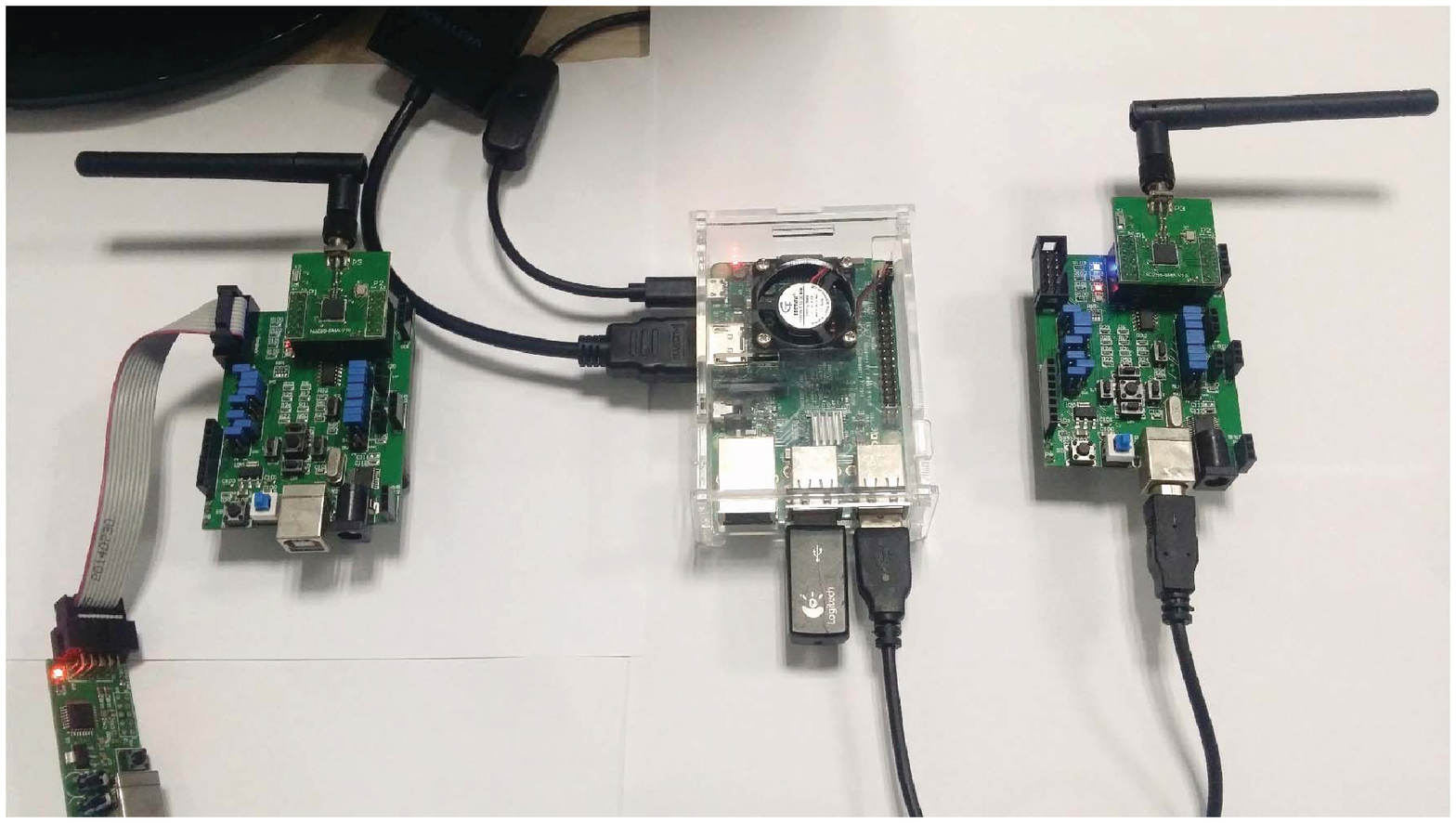}\\
\caption{The energy testing platform}\label{Platform}
\end{center}
\end{figure}

In this subsection, we connect a kind of mote with Raspberry Pi 3B+ via USB interface to simulate the communication between RSU and the data center, as shown in Fig. \ref{Platform}, which is built upon an 8-bit ATmega128L processor with a Chipcon CC2420 radio transceiver. To evaluate the total energy overhead, we calculate the computation energy overhead on Raspberry Pi 3B+ and the communication energy overhead on the mote. By deploying asynchronous counters, we can record the start and end time for the corresponding operation with the precision of 1 millisecond.

\begin{table}
  \centering
  \caption{Mote parameter settings} \label{Parameter Settings}
  \footnotesize
  \tabcolsep 0.05in
  \setlength{\extrarowheight}{0.25cm}
 \begin{tabular}{c|c c}
    \hline
    \quad \quad \raisebox{0.1cm}{Mode} \quad \quad  & \quad \quad  \raisebox{0.1cm}{Current consumption (mA)} \quad \quad \\
    \hline
    Transmitting  &   17.4\\
    Receiving     &   19.7\\
    \hline
 \end{tabular}
\end{table}

Meanwhile, according to the method in \cite{shim2015secure}, we record the size of a message to transmit or receive on the mote. Then, we transfer it into the communication energy consumption via the equation: $W_{r} = V \ast I \ast b \diagup d_{r}$, as shown in Table \ref{Parameter Settings}. Here, ``$W_{r}$'' represents the energy magnitude, ``$V$'' represents the power level of the mote, ``$I$'' represents the current mode of the mote, and ``$b$'' represents the size of a message (in bits), ``$d_{r}$'' represents the the data rate. Moreover, the data rate is 250 kbps, and we assume the power level of the mote is 3.0V. In addition, we can compute the computation energy overhead through the equation: $W = V \ast I \ast t$. Here, ``$W$'' represents the energy magnitude, ``$V$'' represents the power level of the Raspberry Pi 3B+, ``$I$'' represents the current of the Raspberry Pi 3B+, and ``$t$'' represents the time of an operation. Here, we assume that the power level of the Raspberry Pi 3B+ is 5.0V, the current is 1.0A. Then, we obtain the computation energy overhead. Eventually, we analyze the total energy overheads of MDBV as follows:

\subsubsection{RSU}\
\begin{itemize}
    \item \emph{Energy cost on computation}: RSU only make simple aggregation operations, so its energy cost can be ignored;
    \item \emph{Energy cost on communication}: According to \cite{wander2005energy}, we used a packet size of 41 bytes, 32 bytes for the payload and 9 bytes for the header. The header, ensuing a 8-byte preamble, consists of source, destination, length, packet ID, CRC, and a control byte. Thus, we can calculate the RSU's communication consumption for data transmission on the mote as $W_r = 3 \times 17.4 \times (41\times4+20+9+5\times8) \times 8 /250000 = 0.3892032 mJ$;
    \item \emph{Total energy cost}: The total energy overhead for each member node on the mote is $0.3892032mJ$.
\end{itemize}

\subsubsection{Data Center}\
\begin{itemize}
    \item \emph{Energy cost on computation}: The time of verifying a message in MDBV is 113.375 ms, so it requires $5 \times 1 \times 113.375= 566.875 mJ$;
    \item \emph{Energy cost on communication}: The data center receives a message with the length of $n\times S + 2L$ from RSU. Here, $n$ denotes the number of vehicles, and we assume $S = L = 20$ bytes. For simplification of comprehension, we set $n = 1$ and $n = 20$ respectively, and then calculate communication consumption (namely, $W_{rn}$) for receiving message in side of data center, as follows: $W_{r1} = 3 \times 19.7 \times (41\times4+20+9+5\times8) \times 8 /250000 = 0.4406496 mJ$, and $W_{r20} = 3 \times 19.7 \times (41\times16+16+9+8\times17) \times 8 /250000 = 1.5451104 mJ$;
    \item \emph{Total energy cost}: we set $w=\lfloor(20\times n+128)\diagup32 \rfloor$, so the total energy of the data center on the mote is $ 566.875 + (41\times w + 20\times n+137-32\times w + (w+1)\times 8)\times 3 \times 19.7/250000 = 0.0040188w + 0.004728n + 566.909278 mJ$.
\end{itemize}

From the above results, the energy consumption on communication in RSU and data center is both insignificant, compared with the overhead on computation. So, the computation complexity mainly determines the performance of a scheme. Next, assuming $ \left| data \right|= 160 $ bits, we evaluate the total energy overhead in these schemes, as shown in Fig.\ref{total energy}. Consequently, we find that MDBV minimizes energy consumption and achieves the best performance among all the selected schemes.

For all above, our scheme achieves quick data authentication and strengthens the security and survivability of IoV. It is more suitable for realistic IoV scenarios.

\section{Conclusion}
In this paper, we design a CLAS-based monitoring data batch verification scheme for IoV scenarios--MDBV that enhances the survivability and reduces the computation overhead effectively. The scheme is proved to be secure in the random oracle model under the hardness of CDHP. Furthermore, MDBV has the fixed length of the aggregated authentication message with the increasing number of vehicles. Each vehicle can dynamically participate in the system using own information and public system parameter. Moreover, the performance evaluation shows that the computation overhead, communication overhead, and energy cost of the proposed scheme are less than the other relevant schemes. MDBV is more suitable for the survivability of IoV.

\ifCLASSOPTIONcaptionsoff
  \newpage
\fi

\bibliographystyle{IEEEtran}
\bibliography{ms}

\end{document}